\documentclass{aa} 

\usepackage{amssymb}
\usepackage[usenames,dvipsnames]{xcolor}
\usepackage{graphicx}
\usepackage{nicefrac}
\usepackage[rightcaption]{sidecap}
\usepackage{natbib}
\usepackage{txfonts}
\usepackage[normalem]{ulem}
\usepackage{url}
\usepackage{hyperref}
\hypersetup{
    final=true,
    pageanchor=true,
    colorlinks=true,
    breaklinks=true,
    linkcolor=blue,
    citecolor=blue,
    urlcolor=blue,
    pdfpagemode=UseNone,
    pdftitle={Emerging Flux Region},
    pdfauthor={M. Verma},
    pdfsubject={Solar Physics},
    pdfkeywords={Sun: chromosphere, Sun: photosphere, Sun: surface magnetism,
    Sun: sunspots, Techniques: image processing, Methods: data analysis}}
    
\usepackage{pifont}
\usepackage{epstopdf}

\begin{document}

%###############################################################################
%#
%#    TITLE
%#
%###############################################################################

\title{Horizontal flow fields in and around a small active region}
\subtitle{The transition period between flux emergence and decay} 

\author{%
    M.\ Verma\inst{1},
    C.\ Denker\inst{1},   
    H.\ Balthasar\inst{1},
    C.\ Kuckein\inst{1},
    S.J.\ Gonz{\'a}lez Manrique\inst{1,2}, 
    M.\ Sobotka\inst{3},
    N.\ Bello Gonz{\'a}lez\inst{4},\\  
    S.\ Hoch\inst{4},   
    A.\ Diercke\inst{1,2},
    P.\ Kummerow\inst{1,2},   
    T.\ Berkefeld\inst{4}, 
    M.\ Collados\inst{5},
    A.\ Feller\inst{6},
    A.\ Hofmann\inst{1},
    F.\ Kneer\inst{7},\\
    A.\ Lagg\inst{6},
    J.\ L{\"o}hner-B{\"o}ttcher\inst{4},
    H.\ Nicklas\inst{7}, 
    A.\ Pastor Yabar\inst{5, 10},
    R.\ Schlichenmaier\inst{4},
    D.\ Schmidt\inst{8},\\
    W.\ Schmidt\inst{4},
    M.\ Schubert\inst{4},
    M.\ Sigwarth\inst{4},
    S.K.\ Solanki\inst{6, 9},
    D.\ Soltau\inst{4},        
    J.\ Staude\inst{1},\\
    K.G.\ Strassmeier\inst{1},
    R.\ Volkmer\inst{4},
    O.\ von der L{\"u}he\inst{4}, \and
    T.\ Waldmann\inst{4}}

\institute{%
    Leibniz-Institut f{\"u}r Astrophysik Potsdam (AIP),
    An der Sternwarte~16,
    14482 Potsdam,
    Germany,
%    \email{mverma@aip.de}
    \href{mailto:mverma@aip.de}{\textsf{mverma@aip.de}}
\and
    Universit{\"a}t Potsdam,
    Institut f{\"u}r Physik und Astronomie,
    Karl-Liebknecht-Stra{\ss}e 24/25,
    14476 Potsdam-Golm,
    Germany
\and
    Astronomical Institute, Academy of Sciences of the Czech Republic,
    Fri\v{c}ova 298,
    25165 Ond\v{r}ejov, Czech Republic
\and 
    Kiepenheuer-Institut f{\"u}r Sonnenphysik,
    Sch{\"o}neckstr.\ 6,
    79104 Freiburg, Germany
\and
    Instituto de Astrof\'{\i}sica de Canarias,
    C/ V\'{\i}a L\'actea s/n, 
    38205 La Laguna, Tenerife, Spain
\and
    Max-Planck-Institut f{\"u}r Sonnensystemforschung,
    Justus-von-Liebig-Weg 3,
    37077 G{\"o}ttingen, Germany
\and
    Institut f\"ur Astrophysik,
    Georg-August-Universit\"at G\"ottingen,
    Friedrich-Hund-Platz 1,
    37077 G\"ottingen, Germany
\and    
    National Solar Observatory, 
    3010 Coronal Loop
    Sunspot, NM 88349, USA
\and 
    School of Space Research, 
    Kyung Hee University,Yongin, 
    Gyeonggi-Do, 446-701, Republic of Korea
\and    
    Dept. Astrof\'{\i}sica, 
    Universidad de La Laguna, 
    E-38205, La Laguna, Tenerife, Spain}
\authorrunning{Verma et al.}

\date{submitted 25/02/2016, accepted 12/05/2016}

% 5 {} token are mandatory
\abstract
% context heading (optional)
{The solar magnetic field is responsible for all aspects of solar activity. 
Thus, emergence of magnetic flux at the surface is the first manifestation of 
the ensuing solar activity.} 
% aims heading (mandatory)
{Combining high-resolution and synoptic observations aims to provide a 
comprehensive description of flux emergence at photospheric level and of the 
growth process that eventually leads to a mature active region.}
% methods heading (mandatory)
{The small active region NOAA~12118 emerged on 2014 July~17 and was observed one 
day later with the 1.5-meter GREGOR solar telescope on 2014 July~18. 
High-resolution time-series of blue continuum and G-band images acquired in the 
blue imaging channel (BIC) of the GREGOR Fabry-P\'erot Interferometer (GFPI) 
were complemented by synoptic line-of-sight magnetograms and continuum images 
obtained with the Helioseismic and Magnetic Imager (HMI) onboard the Solar 
Dynamics Observatory (SDO). Horizontal proper motions and horizontal plasma 
velocities were computed with local correlation tracking (LCT) and the 
differential affine velocity estimator (DAVE), respectively. Morphological image 
processing was employed to measure the photometric and magnetic area, magnetic 
flux, and the separation profile of the emerging flux region during its 
evolution.}
% results heading (mandatory)
{The computed growth rates for photometric area, magnetic area, and magnetic 
flux are about twice as high as the respective decay rates. The space-time 
diagram using HMI magnetograms of five days provides a comprehensive view of 
growth and decay. It traces a leaf-like structure, which is determined by the 
initial separation of the two polarities, a rapid expansion phase, a time when 
the spread stalls, and a period when the region slowly shrinks again. The 
separation rate of 0.26~km~s$^{-1}$ is highest in the initial stage, and it 
decreases when the separation comes to a halt. Horizontal plasma velocities 
computed at four evolutionary stages indicate a changing pattern of inflows. In 
LCT maps we find persistent flow patterns such as outward motions in the outer 
part of the two major pores, a diverging feature near the trailing pore marking 
the site of upwelling plasma and flux emergence, and low velocities in the 
interior of dark pores. We detected many elongated rapidly expanding granules 
between the two major polarities, with dimensions twice as large as the normal 
granules.}
% conclusion (optional)
{}


\maketitle

%===============================================================================
%    Introduction
%===============================================================================

\section{Introduction}
       
Solar activity is directly related to the Sun's magnetic field. Flux emergence 
affects all atmospheric layers and involves many coupled but diverse physical 
environments. The rise of buoyant magnetic flux tubes from the convection zone 
to the surface and higher solar atmospheric layers eventually leads to the 
formation of sunspots and active regions \citep{Parker1955, Zwaan1978}. The 
rising flux tubes are already concentrated bundles of magnetic field lines when 
they penetrate the photosphere. \citet{Zwaan1985} explained the general dynamics 
of emerging flux regions (EFRs), that is, the separation of opposite polarities 
and coalescence of same-polarity magnetic features by the buoyancy of 
(collapsed) flux tubes.  

% Various studies discussed various observational characteristics of magnetic 
% and velocity fields in EFR.
Typically, EFRs consist of many pores and protopores in various stages of growth 
or decay with separating polarities and sometimes with strong photospheric shear 
flows \citep{Brants1985a}. The properties and dynamics of small-scale magnetic 
structures within an EFR were thoroughly examined by \citet{Strous1996} and 
\citet{Strous1999}. They found thread-like concentrations of magnetic flux in 
facular elements within an EFR. Flux emergence occurred throughout the region, 
which was governed by the separation of opposite polarities with a velocity of 
about 1~km~s$^{-1}$.

% magnetic field observations
The vector magnetic field in and around an EFR was studied by \citet{Lites1998}. 
Their observations were in accordance with earlier ideas of emerging bipolar 
flux, that is, the buoyant flux tubes transport mass from the photosphere to the 
chromosphere, and subsequently material drains down to footpoints along arched 
magnetic loops. These authors concluded that the formation of pores or sunspots 
by coalescence of magnetic flux has its origin in the emergence of subsurface 
structures. This picture was confirmed and further elaborated by 
\citet{Solanki2003} and \citet{Xu2010}, who determined the magnetic and velocity 
structure of EFRs in both the photosphere and the chromosphere from the 
Si\,\textsc{i} 10827 and the He\,\textsc{i} 10830\AA\ lines. The strength of the 
magnetic field in young EFRs is weak \mbox{($<500$~G)}. The magnetic field 
becomes stronger \mbox{($>500$~G)} and vertical within half a day after 
emergence. Additionally, the filling factor decreases from about 80\% to 40\% 
\citep{Kubo2003}. Spectropolarimetric observations presented by 
\citet{Watanabe2008} of an EFR in the Fe\,\textsc{i} 630.2~nm spectral region 
together with H$\alpha$ filtergrams showed that magnetic flux emergence was 
associated with the appearance of Ellerman bombs. The complex velocity and 
magnetic fields of EFRs were the focus of a study by \citet{Luoni2011}. They 
found global magnetic twist during the emergence phase in a sample of 40 active 
regions using elongated `magnetic tongues' \citep{LopezFuentes2000} as a proxy 
for the magnetic helicity.

% Horizontal and LOS flows observations
The interaction of rising flux with photospheric plasma during and after 
emergence is one aspect of the EFR evolution. The separation of opposite 
polarities is often observed \citep{Brants1985a}, and horizontal flows play a 
significant role in the growth of EFRs. \citet{Otsuji2011} inferred horizontal 
velocities, modes of emergence, and total magnetic flux of 101 flux emergence 
events using Hinode data. They found converging as well as diverging flows in 
various emerging events. The computed growth of magnetic flux, separation of two 
polarities, and velocity of separation followed power laws. \citet{Otsuji2011} 
obtained a physical description of flux emergence, where buoyant emerging 
magnetic fields evolve in balance with the surrounding turbulent atmosphere. 
\citet{Toriumi2012} detected horizontal divergent flows prior to flux emergence 
using Helioseismic and Magnetic Imager (HMI) dopplergrams and magnetograms. 
These flows are caused by plasma escaping horizontally from the rising flux 
announcing a newly emerging active region. In addition, horizontal flows 
separating positive and negative polarities are higher in the beginning but 
decrease over time. Isolated flux emergence, that is, outside of pre-existing 
magnetic fields, was analyzed by \citet{Centeno2012}, also using Solar Dynamics 
Observatory (SDO) observations. She noted that EFRs start as simple magnetic 
bipolar regions growing by merging of various small magnetic features along with 
systematic downflows at the footpoints. 

% Theoretical aspect
The magnetohydrodynamic (MHD) simulations of \citet{Cheung2010} reproduced 
observational properties of flux emergence such as elongated granules, mixed 
polarity patterns in the emerging flux region, and pore formation. They 
kinematically advected a magnetic semi-torus upward from a depth of 7.5~Mm to 
the photosphere. \citet{Stein2011} simulated flux emergence using the rise and 
evolution of untwisted horizontal flux with the same entropy as the 
nonmagnetized ascending plasma, which was carried by upflows from a depth of 
20~Mm. \citet{Rempel2014} expanded the work of \citet{Cheung2010} and included 
the emergence and evolution of buoyant flux tubes. In their work, the flux is 
able to emerge to the photosphere even in the absence of twist. 

% What we are doing in this study? 
Collectively, various observational studies and numerical simulations present a 
physical description of flux emergence on the solar surface. However, 
high-resolution observations of the solar surface are needed to enhance our 
understanding of the complex interaction between horizontal flows and magnetic 
fields during the process of flux emergence. In the current study, we present 
the first analysis of observations made with the GREGOR solar telescope, taken 
during a 50-day early science campaign in 2014 July\,--\,August.

%===============================================================================
%    Observations
%===============================================================================

\section{Observations and data reduction}

%-------------------------------------------------------------------------------
%    Figure 2: Overview figure from SDO
%-------------------------------------------------------------------------------
\begin{figure}
\includegraphics[width=\columnwidth]{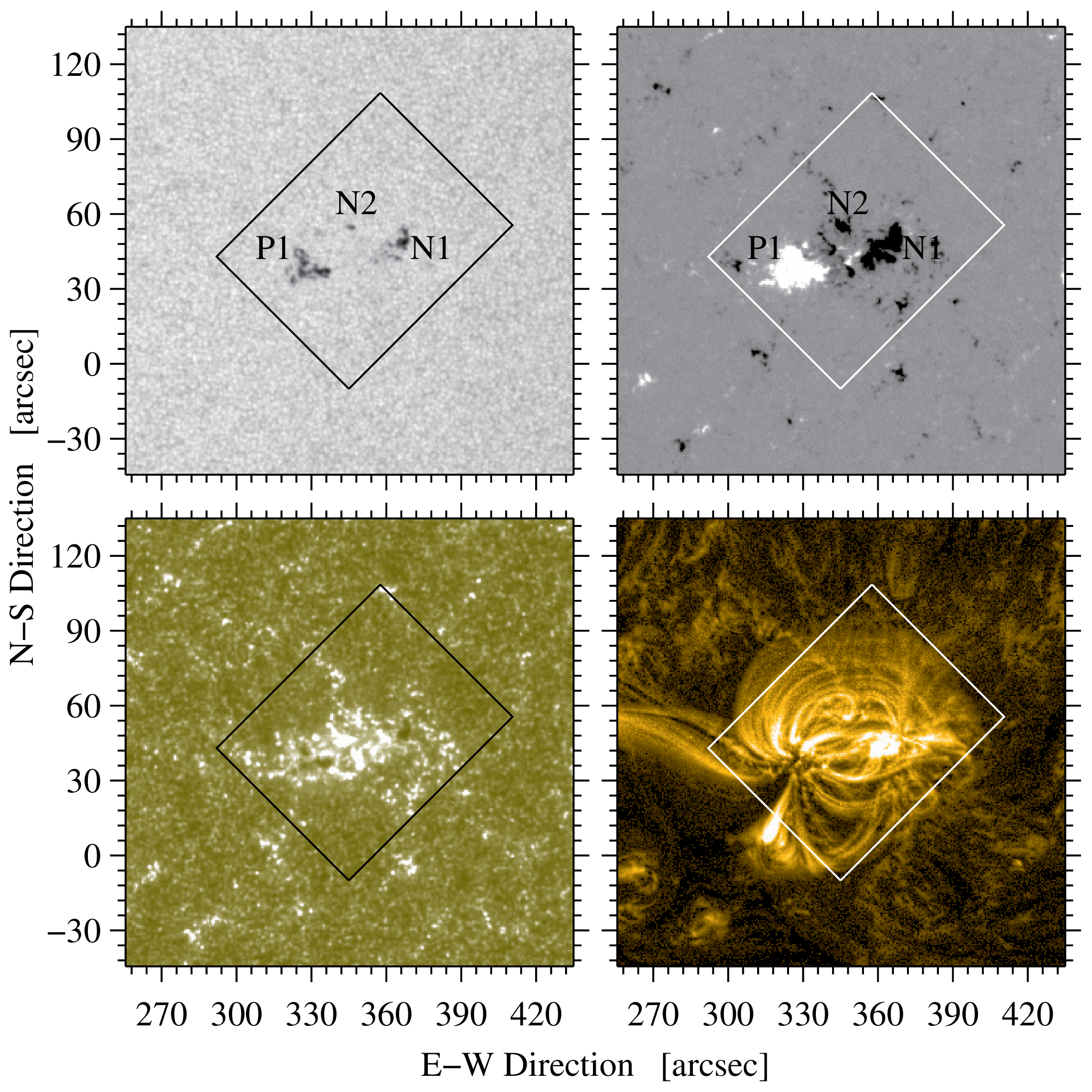}
\caption{Overview of active region NOAA 12118: HMI continuum image
    (\textit{top left}), HMI magnetogram (\textit{top right}), AIA
    $\lambda$160~nm image (\textit{bottom left}), and AIA Fe\,\textsc{ix}
    $\lambda$17.1~nm image (\textit{bottom right}) observed on 
    2014 July~18 at 08:00~UT. A bipolar EFR appeared on
    2014 July~17 to the west of the central meridian in the northern
    hemisphere. Observations of active region NOAA~12118 started at 08:00~UT and
    ended at 10:40~UT on 2014 July~18. The rectangle is the FOV covered by
    GREGOR high-spatial resolution data. The two major pores and one smaller
    pore are marked as \textsf{P1}, \textsf{N1}, and \textsf{N2} to aid 
    discussions in Sect.~\ref{SEC03.3}}
\label{FIG02}
\end{figure}
%-------------------------------------------------------------------------------

\subsection{Observations\label{SEC02.1}}

% About GREGOR
Newly emerging flux was observed with GREGOR Fabry-P\'erot Interferometer 
(GFPI) / blue imaging channel (BIC) \citep{Denker2010b, Puschmann2012} at the 
1.5-meter GREGOR solar telescope \citep{Denker2012, Schmidt2012} on 2014 July~18 
starting at 07:56~UT. The GFPI was originally developed by the Institute for 
Astrophysics in G\"ottingen \citep{Bendlin1992, Puschmann2006, Bello2008}. The 
emerging bipolar region was labeled active region NOAA~12118 on the next day. 
The image quality was improved in real-time by the GREGOR adaptive pptics 
system \citep[GAOS,][]{Berkefeld2010, Berkefeld2012}, which locked on a small, 
circular pore in the center of the active region located in between the pair of 
clustered, opposite-polarity pores.

%-------------------------------------------------------------------------------
%    Figure 1: Blue continuum image from GREGOR
%-------------------------------------------------------------------------------
\begin{figure*}
\includegraphics[width=0.98\textwidth]{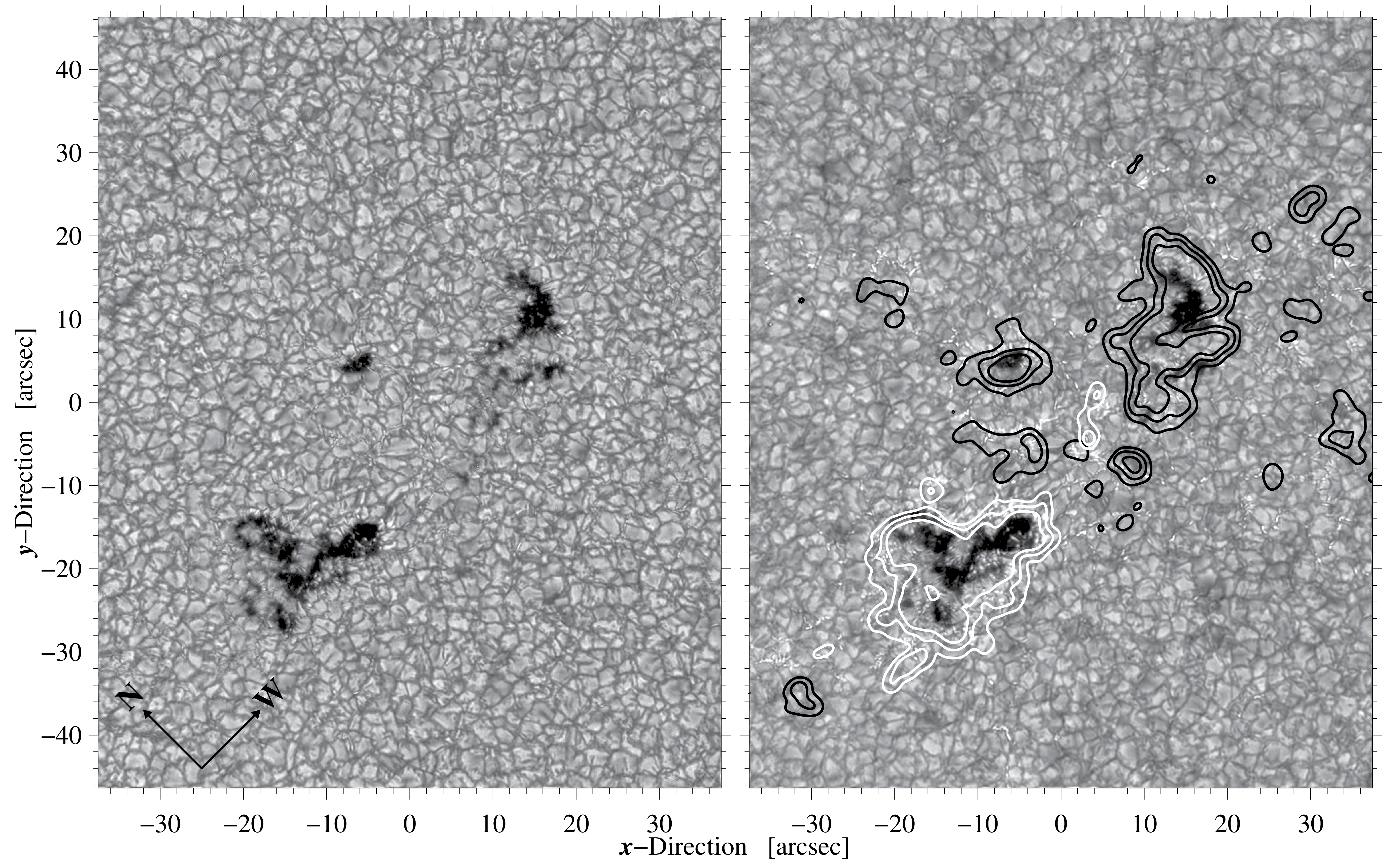}
\caption{Blue continuum (\textit{left}) and G-band (\textit{right}) images of 
the time-series observed on 2014 July~18 at 08:46~UT and 08:43~UT, respectively. 
Images were restored with KISIP. The two images are scaled between 0.4 and 1. 
The black and white contours superposed on the G-band image are created using a 
HMI magnetogram displayed at levels $\pm$100, $\pm$250, and $\pm$500 G. The 
north and west directions are indicated by the arrows in the lower left corner.}
\label{FIG01}
\end{figure*}
%-------------------------------------------------------------------------------

% What settings for chosen for observations
The blue continuum observations $\lambda$450.6~nm were split into four 30-minute 
time-series with 60 image sequences containing 80 images each with an exposure 
time of 4~ms, which are appropriate for image restoration. The four time-series 
started at 07:56~UT, 08:36~UT, 09:13~UT, and 10:09~UT, respectively. The image 
scale of 0.035~\arcsec~pixel$^{-1}$ for the 2160$\times$2672-pixel images leads 
to a field-of-view (FOV) of $75^{\prime\prime} \times 93^{\prime\prime}$ 
covering the entire region and the complex flow field associated with the 
growing and separating regions of opposite magnetic polarity. The cadence of 
30~s for the time-series of restored blue continuum images is ideally suited for 
horizontal flow measurements \citep{Verma2011}. Auxiliary G-band images 
$\lambda$430.7~nm were also taken but camera problems prevented us from taking 
continuous time-series. However, in some cases the time-series were of 
sufficient duration to allow us tracking G-band brightenings, which are often 
associated with small-scale magnetic flux elements \citep{Uitenbroek2006a, 
Leenaarts2006}. The best images of the restored blue continuum and G-band 
time-series are featured in Fig.~\ref{FIG01}. The black arrows show the 
orientation of the BIC FOV on the solar disk at the given time. However, the 
altitude-azimuth mount of the telescope \citep{Volkmer2012} introduces an image 
rotation, that has to be taken into account when analyzing time-series.

% About region observed
The observed active region NOAA~12118 started emerging on July~17 around 
15:00~UT at N7$^\circ$  W41$^\circ$ (disk-center coordinates) and was 
categorized as an $\alpha\beta$-group according to the Hale classification 
\citep{Hale1919}. At the time of the GREGOR observations on July~18, the region 
attained its maximum size. By July~19 the region started to fade out, and by 
July~22 it had decayed significantly before rotating off the visible disk. 
Although C-class flare activity was predicted \citep{Gallagher2002a}, the region 
did not produce any noteworthy flare.

% SDO data
The temporal evolution of the magnetic flux is based on full-disk continuum 
images and line-of-sight (LOS) magnetograms of HMI \citep{Schou2012, 
Couvidat2012, Wachter2012} onboard SDO \citep{Scherrer2012}. To appraise the 
active region in higher atmospheric layers, we used an Fe\,\textsc{ix} image 
$\lambda$17.1~nm of the Atmospheric Imaging Assembly \citep[AIA,][]{Lemen2012}. 
We improved the contrast of this image using noise adaptive fuzzy equalization 
\citep[NAFE,][]{Druckmueller2013}. Figure~\ref{FIG02} gives an overview of 
active region NOAA~12118. The black and white rectangular boxes represent the 
FOV observed by GREGOR/BIC. 

%-------------------------------------------------------------------------------
%    Figure 3: Region time evolution from SDO continuum and Magnetogram
%-------------------------------------------------------------------------------
\begin{figure*}
\includegraphics[width=\textwidth]{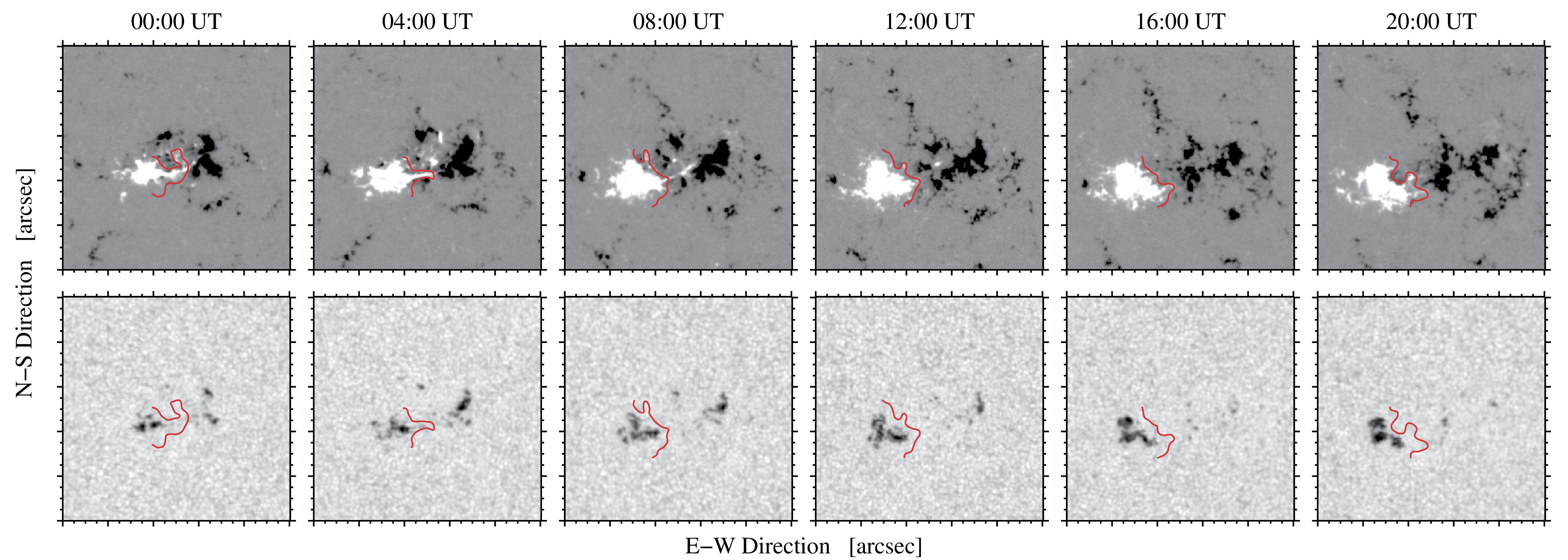}
\caption{Pores belonging to active region NOAA~12118 were visible on the solar 
    surface for about four days (2014 July 17\,--\,20). Cutouts of HMI 
    magnetograms and continuum images ($100\arcsec \times 100 \arcsec$) show 
    the temporal evolution of the region at four-hour intervals centered on the 
    time of the GREGOR observations, that is, 08:00~UT on July~18. The 
    magnetograms are displayed between $\pm$250~G. The red line marks the 
    magnetic neutral line at the respective time.}
\label{FIG03}
\end{figure*}
%-------------------------------------------------------------------------------

\subsection{Data reduction\label{SEC02.2}}

% Speckle reconstruction: KIS Woeger code
The blue continuum and G-band images were restored using the Kiepenheuer 
Institute speckle interferometry package \citep[KISIP,][]{Woeger2008b, 
Woeger2008a}. In this code the image restoration is performed in the Fourier 
domain. The code separates the Fourier phases of the images from their 
amplitudes. Fourier phases can be retrieved by two different phase 
reconstruction algorithms: by the Knox-Thompson \citep{Knox1974} and 
triple-correlation \citep{Weigelt1983}, whereas Fourier amplitudes are 
reconstructed independently \citep{Labeyrie1970}. Figure~\ref{FIG01} shows 
images reconstructed with the triple-correlation technique. The image contrast, 
degraded by scattered light, was then restored using the method described in 
\citet{Bello2008}.

% Local correlation tracking
We applied LCT \citep[for details see][]{Verma2011, Verma2013} to  the blue 
continuum image sequences to compute horizontal flows. The restored time-series 
is aligned  (image rotation and image displacement) with sub-pixel accuracy with 
respect to the average image of the time-series, and the signature of 
five-minute oscillation was removed using a subsonic Fourier filter with a 
cut-off velocity corresponding to the photospheric sound speed. The LCT 
technique computes cross-correlations over $48 \times 48$-pixel image tiles with 
a Gaussian kernel having a FWHM = 1200~km corresponding to the typical size of a 
granule. The time cadence was $\Delta t = 60$~s, and the flow maps were averaged 
over $\Delta T = 30$~min.

% SDO continuum images
We chose from the SDO/HMI database one continuum image and LOS magnetogram with 
$4096 \times 4096$ pixels every hour for the period from 2014 July 17\,--\,22, 
which is a total of 120 full-disk images and magnetograms. The image scale is in 
both cases about 0.5\arcsec\ pixel$^{-1}$. The continuum images were corrected 
for limb-darkening to produce contrast-enhanced images \citep[see 
e.g.,][]{Denker1999a}. The magnetic and photometric evolution of the active 
region on July~18 is shown in Fig.~\ref{FIG03}. The magnetic flux values were 
corrected for projection effects assuming that all flux tubes are perpendicular 
to the solar surface.

% DAVE
We selected four different time sequences of HMI LOS magnetograms with a 
cadence of 45~s to study the magnetic field evolution of the region and to 
compute horizontal plasma velocities. The selected time sequences were from 
15:30\,--\,17:30~UT on July~17, from 08:00\,--\,10:00~UT on July~18, from 
00:00\,--\,02:00~UT on July~19, and from 21:30\,--\,23:30~UT on July~19. The 
basic calibration included dark and flat-field corrections, elimination of 
spikes, and removal of geometric foreshortening. In the next step, we applied 
DAVE \citep{Schuck2005, Schuck2006} to the magnetogram sequence to retrieve the 
horizontal plasma velocities. The underlying physics is based on magnetic 
induction equation, and the implementation of DAVE implies an affine velocity 
profile. The horizontal flows were estimated using temporal and spatial 
derivatives of the magnetic field. We applied the Scharr operator  
\citep{Scharr2007} for the spatial derivatives and a five-point stencil with a 
time difference of 15~min between neighboring magnetograms for the temporal 
derivative. The horizontal plasma velocities depicted in Fig.~\ref{FIG06} are 
averages of 80 individual DAVE maps covering one hour. However, the temporal 
derivatives add 30~min before and after this time interval so that the 
horizontal plasma velocities contain information from a two-hour period. We 
employed DAVE because it provides more accurate results than applying LCT to 
magnetograms \citep{Schuck2005, Schuck2006}.

%===============================================================================
%    Results
%===============================================================================

\section{Results}

In the following sections, we present the main results of this study. Starting 
with the temporal evolution of the region size and magnetic flux contents, we 
discuss horizontal plasma velocities computed using DAVE, and we study changes 
in horizontal proper motions.

%-------------------------------------------------------------------------------
%    Figure 4: Magnetic field evolution from SDO magnetogram
%-------------------------------------------------------------------------------
\begin{figure*}
\includegraphics[width=\columnwidth]{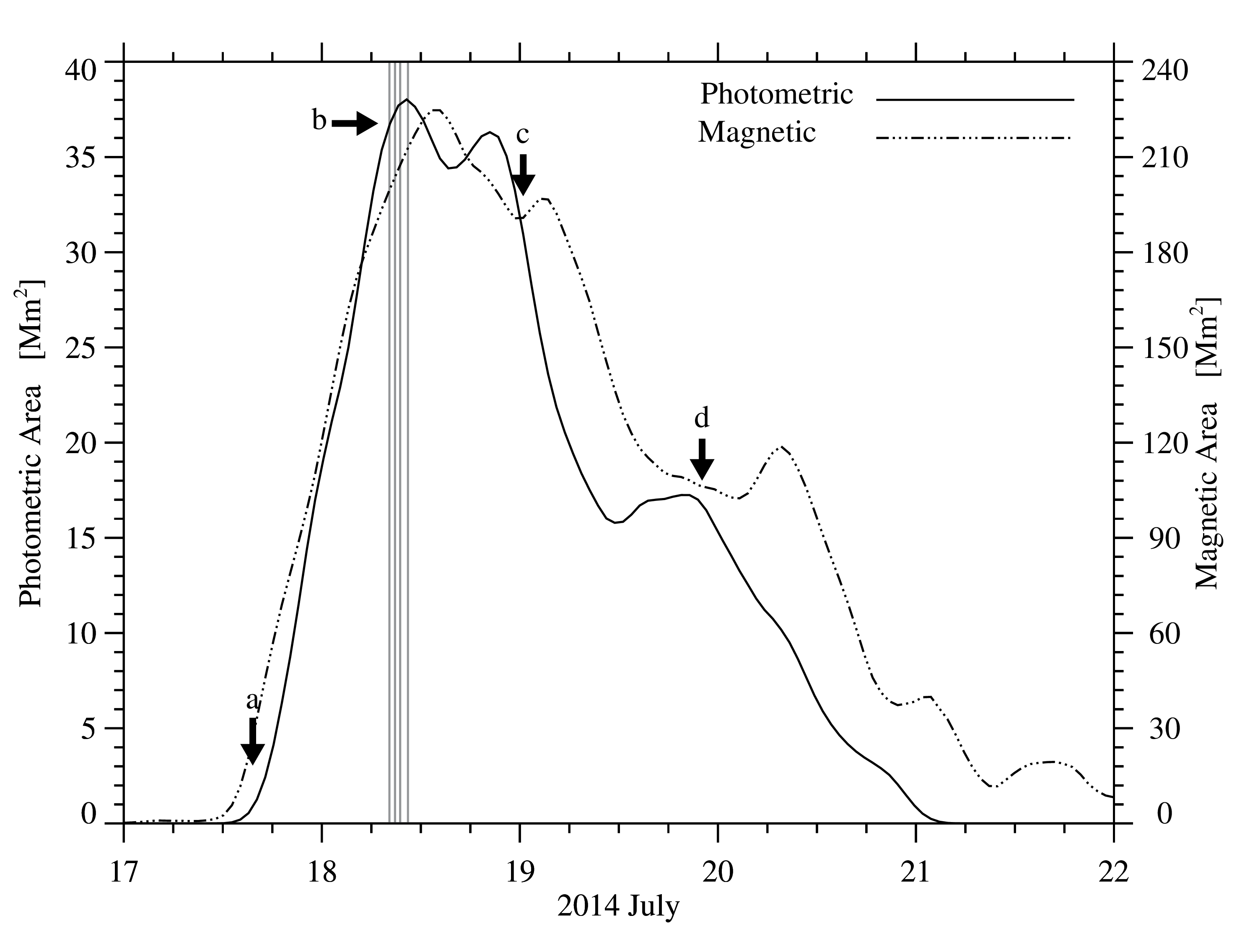}
\includegraphics[width=\columnwidth]{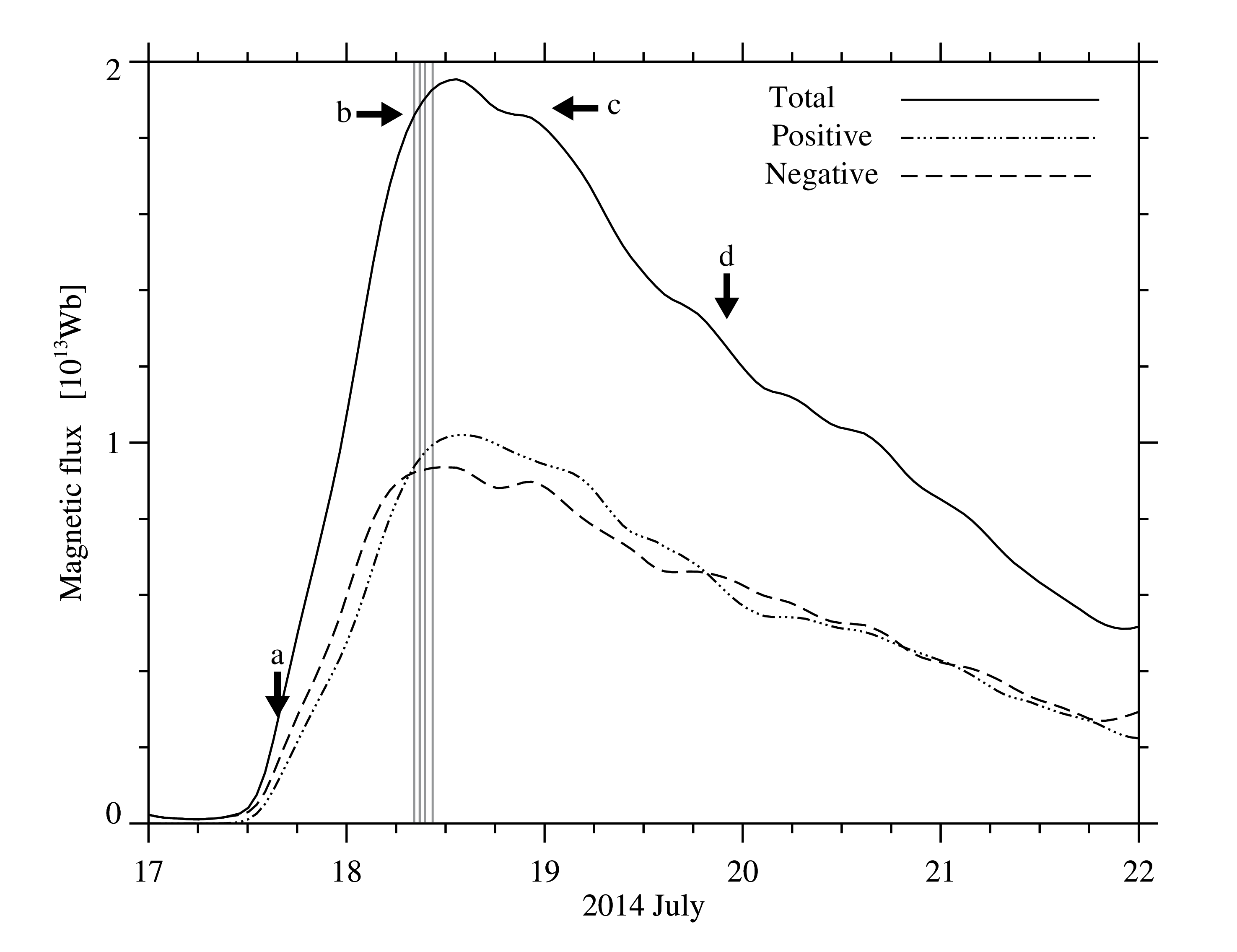}
\caption{Temporal evolution of the area (\textit{left}) and magnetic flux 
    (\textit{right}) in active region NOAA~12118 during its disk passage. The 
    solid and dash-dotted lines in the left panel show the photometric and 
    magnetic area, respectively. The solid, dash-dotted, and dashed lines in 
    the right panel correspond to the total, positive, and negative magnetic 
    flux, respectively.  The four vertical lines indicate the time of the BIC 
    time-series. The labels a to \-d indicate different stages of active 
    region evolution, which are explained in Sect.~\ref{SEC03.2}.}
\label{FIG04}
\end{figure*}
%-------------------------------------------------------------------------------

\subsection{Photometric and magnetic evolution}

% \citep{Bernasconi2002}: moving dipolar features in EFR \\
The emerging flux region appeared on the solar disk  at about 15:00~UT on 
July~17. The EFR surfaced in a quiet-Sun region without any major active region 
in the vicinity. The bipolar region started out with two pores. The two major 
pores were eventually joined by smaller pores of both polarities. The region was 
fully developed by the time of the GREGOR observations. Both main pores were 
increasing in area, and the separation between them grew as well. Snapshots of 
the region's evolution are shown for July~18 in Fig.~\ref{FIG03}. The HMI 
continuum images and LOS magnetograms are displayed every four hours. We grouped 
these snapshots into three phases: (1) growth phase 00:00\,--\,04:00~UT, (2) 
maximum phase 08:00\,--\,12:00~UT, and (3) decay phase 16:00\,--\,20:00~UT. In 
phase (1) the leading part with negative polarity consists of two smaller pores, 
and another small pore of the same polarity is developing in the center of the 
FOV. The trailing part is more compact at 00:00~UT, in contrast to the typical 
appearance of active regions. By phase (2), both leading and trailing parts have 
grown. The leading polarity started to dissolve at 12:00~UT, as is evident in 
the continuum image, while the trailing part continues to grow in area. After 
four hours in phase (3), any traces of the leading polarity have vanished in the 
continuum images, but the trailing part continues to increase in area even until 
20:00~UT.

After July~18, the region decayed significantly. Over the course of their 
evolution, the pores never developed a penumbra, that is, the active region 
never evolved into a group with sunspots. As already mentioned, the leading 
pores dissolved first, while the trailing pores became more compact. The second 
phase of flux emergence after the appearance of two polarities is the 
coalescence of small-scale magnetic flux elements \citep{Strous1999, 
Centeno2012}, which we also observed in this region. Although the region had two 
main pores of opposite polarity, the magnetic neutral line was initially 
complex, but became simpler with time. Initially, a tongue of positive polarity 
extended into the negative-polarity territory. Then, the central part of the 
active region exhibited mixed polarity. Finally, the opposite polarities were 
clearly separated. No arch filament structure was present in H$\alpha$ full-disk 
images obtained with the Chromospheric Telescope 
\citep[ChroTel,][]{Kentischer2008, Bethge2011}, instead, the region contained 
bright plage with only a single filament connecting the two opposite polarities.

We used the contrast-enhanced HMI full-disk continuum images and LOS 
magnetograms to follow the photometric and magnetic evolution of the region over 
five days. The overall evolution of area and flux contained in the active region 
are shown in Fig.~\ref{FIG04}. The dark pores were identified with the help of 
intensity thresholding and morphological image processing. We used a fixed 
intensity threshold of $I = 0.8I_{0}$, where $I_{0}$ refers to the normalized 
quiet-Sun intensity. We took the measured magnetic field strength at face value 
and only carried out a geometrical correction to yield the proper average values 
of the magnetic flux. We use the term magnetic field strength throughout the 
text to refer to the longitudinal flux. To compute the temporal evolution of the 
magnetic flux, we created a binary template that only contains pixels above or 
below $\pm$50 Gauss in the HMI magnetograms. Morphological erosion with a kernel 
of 1 Mm was applied to the template three times to eliminate small isolated flux 
concentrations. Finally, we used morphological dilation with a kernel of 5 Mm to 
include the strong magnetic fields in the immediate neighborhood of the active 
region. This corresponds to a modified morphological opening operation.

%---- Table 1 ------------------------------------------------------------------
%\input{table1.tex}
\begin{table}[t]
\begin{center}
\caption{Growth and decay rates for area and magnetic flux.}\medskip
\small
\label{TAB01}
\begin{tabular}{lcc}
\hline\hline
Area & Growth              & Decay \rule[-2mm]{0mm}{6mm}\\
     & [Mm$^2$~day$^{-1}$] & [Mm$^2$~day$^{-1}$]\rule[-2mm]{0mm}{3mm}\\
\hline
Photometric & $46.6$  & $23.5$ \rule{0mm}{4mm}\\
            &         & $13.9$ \rule{0mm}{4mm}\\
Magnetic    & $226.4$ & $99.2$ \rule{0mm}{4mm}\\
            &         & $97.3$ \rule{0mm}{4mm}\\
\hline
Magnetic flux & Growth          & Decay \rule[-2mm]{0mm}{6mm}\\
              & [Wb~day$^{-1}$] & [Wb~day$^{-1}$]\rule[-2mm]{0mm}{3mm}\\
\hline              
Total    & $2.02$ & $0.44$ \rule{0mm}{4mm}\\
Positive & $1.02$ & $0.24$ \rule{0mm}{4mm}\\
Negative & $1.01$ & $0.20$ \rule{0mm}{4mm}\\
\hline 
\end{tabular}
\end{center}
\end{table}
%-------------------------------------------------------------------------------

%-------------------------------------------------------------------------------
%    Figure 5 : Space-time slice figure
%-------------------------------------------------------------------------------
\begin{figure}
\includegraphics[width=\columnwidth]{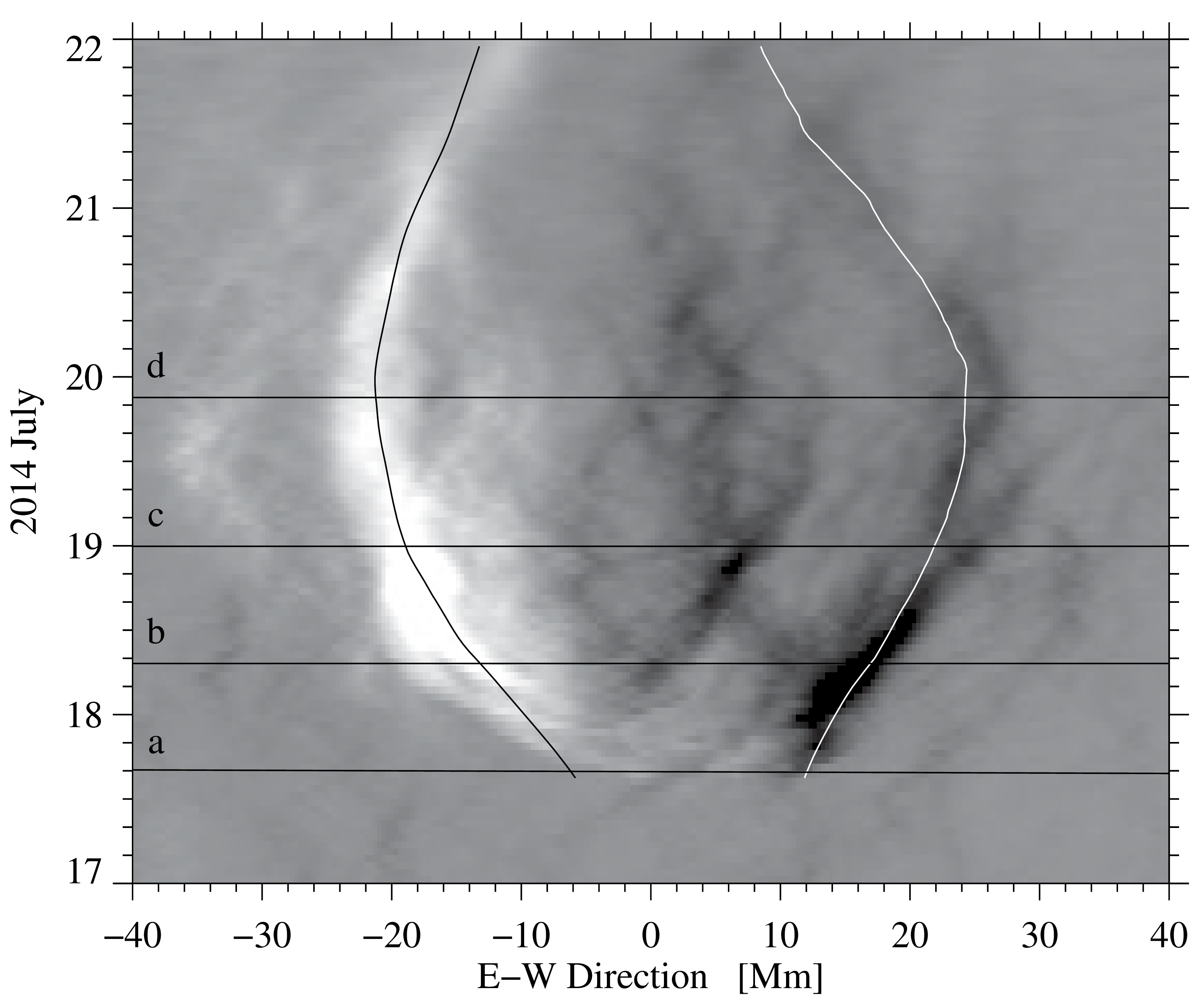}
\caption{Space-time diagram collating five days (July 17\,--\,21) of HMI
    magnetograms. The magnetic field strength is scaled between $\pm 100$~G. 
    The white and black curves trace the location of the extrema of negative 
    and positive polarities over five days. The black horizontal lines indicate 
    the four stages discussed in Sect.~\ref{SEC03.2}}
\label{FIG05}
\end{figure}
%-------------------------------------------------------------------------------

%-------------------------------------------------------------------------------
%    Figure 6: DAVE arrows over SDO LOS-Magnetogram
%-------------------------------------------------------------------------------
\begin{figure*}[t]
\includegraphics[width=\textwidth]{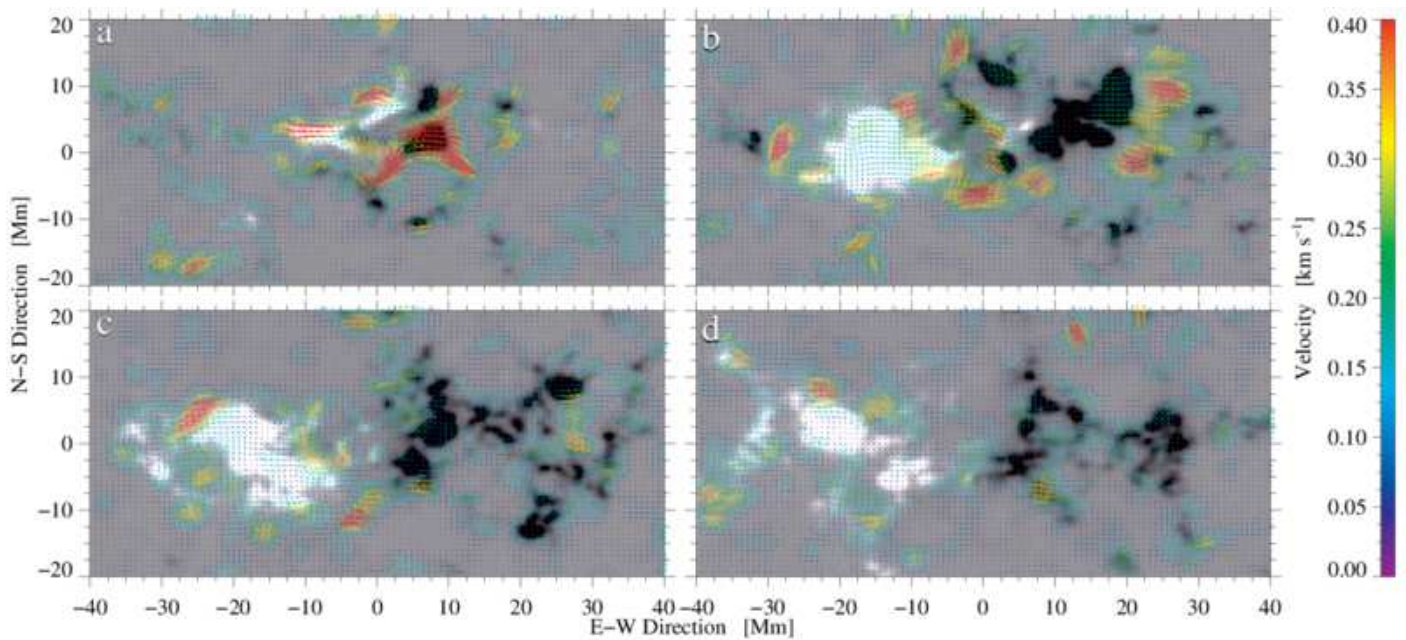}
\caption{Horizontal plasma velocities measured with DAVE for two-hour 
     time-series of 45-second HMI magnetograms depicting the four 
     evolutionary stages of the region: (a) 15:30\,--\,17:30~UT on July~17, (b) 
     08:00\,--\,10:00~UT on July~18, (c) 00:00\,--\,02:00~UT on July~19, and 
     (d) 21:30\,--\,23:30~UT on July~19. Color-coded vectors (best seen when
     zooming in in the on-line version, and red indicates a flow speed in 
     excess of 0.4~km~s$^{-1}$) are superposed onto a time-averaged magnetogram 
     covering the respective time-series.}
\label{FIG06}
\end{figure*}
%-------------------------------------------------------------------------------

We used linear regression to compute the growth and decay rates. A linear fit to 
the HMI continuum data is appropriate because there is no indication for a 
parabolic (or any other non-linear) growth or decay law. There are two peaks on 
July~18, best visible in the photometric area record, but also weakly seen in 
the total flux, one around the time of the GREGOR observations 10:28~UT and 
another at 20:36~UT. The second peak arises because of the still growing, 
trailing part of the region. However, at the same time, the leading part had 
already entirely dissolved. We computed only one growth rate, namely from the 
start of the growth phase to just before the first peak. The photometric growth 
rate for the area is 46.6~Mm$^2$~day$^{-1}$. The decay of the photometric area 
of the region also occurs in two stages. The two local maxima during the decay 
phase lead to two decay rates. First the region decays rapidly at a rate of 
23.5~Mm$^2$~day$^{-1}$, which is comparable to the growth rate. Starting on 
July~20, it decays more slowly at 13.9~Mm$^2$~day$^{-1}$, and by July~21 the 
photometric imprint of the region has completely vanished. Typical decay rates 
for sunspots are in the range from 10\,--\,180~Mm$^{2}$~day$^{-1}$ , as 
discussed in many studies \citep[e.g.,][]{Bumba1963, Moreno-Insertis1988, 
Pillet1993, Hathaway2008}, and in the observed active region with pores the 
decay rate is still within this range. \citet{Rucklidge1995} observed an overlap 
between the radii of large pores and small spots. This overlap is a consequence 
of a convective mode that sets in suddenly and rapidly when the inclination of 
the photospheric magnetic field exceeds some critical value. As a result, a 
filamentary penumbra is formed; this process is not observed in our dataset.

Additionally, we computed the magnetic area using a binary mask based on HMI LOS 
magnetograms. The magnetic area is 4\,--\,5 times larger than the photometric 
area (note the two different scales in the left panel of Fig.~\ref{FIG04}). 
However, it follows the same trend as the photometric area during the decay 
phase, which again has two stages, but shifted by about half a day. The growth 
rate is 226.4~Mm$^2$~day$^{-1}$, which is five times the photometric growth 
rate. The decay rates are very similar 99.2~Mm$^2$~day$^{-1}$ and 
97.3~Mm$^2$~day$^{-1}$ for the first and second half, respectively, in contrast 
to the photometric decay rates. The growth rate is faster than the corresponding 
decay rates \citep[e.g.,][]{Verma2012a}. As the photometric area approaches 
zero, magnetic structures become smaller: first dark pores shrink, then they 
developed into dark magnetic knots \citep{Beckers1968} with sizes smaller than a 
granule, before dispersing as small-scale bright points \citep{Berger1995} in 
regions of abnormal granulation \citep{deBoer1992}. The spatial resolution of 
HMI magnetograms is insufficient to detect magnetic knots or even bright points, 
that is, the magnetic filling factor in combination with the spatial resolution 
affects the time when the photometric decay rate reaches zero. Magnetic flux 
will still be present at this time, so that afterward the development of the 
active region is governed by the magnetic decay rate.

%-------------------------------------------------------------------------------
%    Figure 7: LCT arrows over Blue continuum GREGOR Image
%-------------------------------------------------------------------------------
\begin{figure*}
\includegraphics[width=\textwidth]{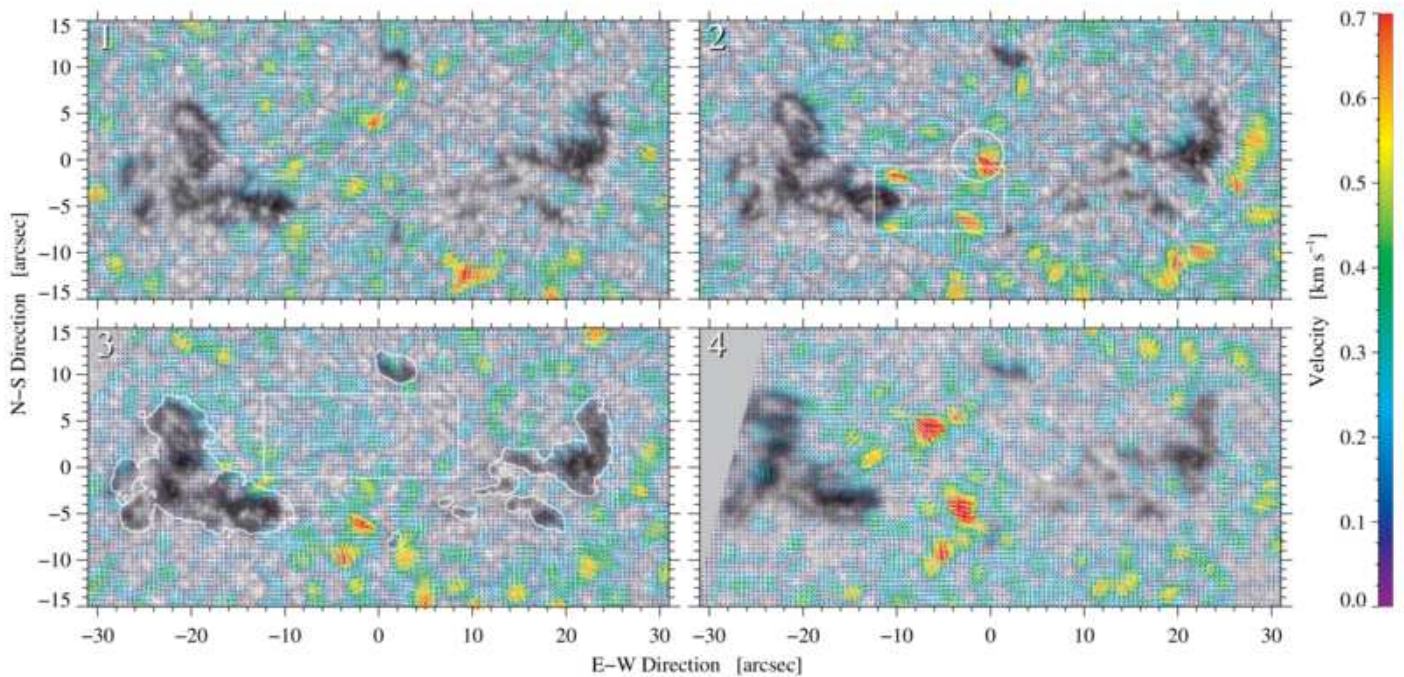}
\caption{Horizontal flow field around active region NOAA~12118 computed 
    using GREGOR blue continuum images. Color-coded vectors (best seen when
     zooming in in the on-line version and red indicates a 
    flow speed in excess of 0.7~km~s$^{-1}$) are superposed onto the first 
    image of the respective time-series. The region is rotated for better 
    display such that its longest axis is horizontal. The white circle in 
    panel~(2) indicates a large, elongated, rapidly expanding granule (see 
    Fig.~\ref{FIG10}). The white rectangle in panel~(2) marks the diverging 
    feature present in all panels but it is strongest in panel~(2). The white 
    contour lines and rectangle in panel~(3) refer to the regions that are 
    used to characterize the horizontal flow speed for various features (see 
    Fig.~\ref{FIG08} and Table~\ref{TAB03}).}
\label{FIG07}
\end{figure*}
%-------------------------------------------------------------------------------

The growth rates of the magnetic flux are 2.02, 1.02, and 1.01 $\times 
10^{13}$~Wb~day$^{-1}$ for the total, positive, and negative flux, respectively. 
The positive and negative flux showed a monotonous rise until the middle of 
July~18. The growth rates for positive and negative flux are virtually 
identical. The flux growth rates agree with the result presented by 
\citet{Otsuji2011}. The decay rates of the magnetic flux are 4\,--\,5 times 
lower than the growth rates and amounted to 0.44, 0.24, and 0.20 $\times 
10^{13}$~Wb~day$^{-1}$ for the total, positive, and negative flux, respectively. 
The decay rates for this small region with pores are in agreement with 
\citet{Kubo2008a}, who found a decay rate of about $ 0.67 \times 10^{13}$~Wb 
day$^{-1}$ in a decaying active region. We list all the growth and decay 
rates in Table~\ref{TAB01}.

% \textcolor{Black}{More details-- Separation rates of the two polarities}
An important evolutionary feature describing the emergence of an active region 
is the rate at which the two polarities separate. To gain insight into the 
overall motion of polarities, we created a space-time diagram (Fig.~\ref{FIG05}) 
encompassing the five days of HMI magnetograms. Since the two major polarities 
are horizontally aligned along the E-W direction, the images corrected for 
geometrically foreshortening are averaged along the vertical axis, that is, 
along the N-S direction. The complete evolution of the active region is thus 
represented in just one diagram. We computed the maximum and minimum of the 
positive and negative polarities for each point in time and represent them as 
black and white curves, which create a leaf-like structure.

%---- Table 2 ------------------------------------------------------------------
% \input{table2.tex}
\begin{table}[t]
\begin{center}
\caption{Parameters describing the horizontal plasma velocities shown in the four
panels of Fig.~\ref{FIG06}.}\medskip
\small
\label{TAB02}
\begin{tabular}{lccccc}
\hline\hline
          & $\bar{v}$ & $v_\mathrm{med}$ & $v_{10}$ & $v_\mathrm{max}$ & $\sigma_{v}$\rule[-2mm]{0mm}{6mm}\\
          & [km~s$^{-1}$] & [km~s$^{-1}$] & [km~s$^{-1}$] & [km~s$^{-1}$] & [km~s$^{-1}$]\rule[-2mm]{0mm}{3mm}\\
\hline
Panel `a' & $0.10$ & $0.07$ & $0.23$ & $0.77$ & $0.10$ \rule{0mm}{4mm}\\
Panel `b' & $0.12$ & $0.10$ & $0.26$ & $0.55$ & $0.10$ \rule{0mm}{4mm}\\
Panel `c' & $0.11$ & $0.08$ & $0.20$ & $0.51$ & $0.07$ \rule{0mm}{4mm}\\
Panel `d' & $0.09$ & $0.07$ & $0.17$ & $0.45$ & $0.06$ \rule{0mm}{4mm}\\
\hline
\end{tabular}
\footnotesize \hspace{1mm}
\parbox{80mm}{% \vspace{-1mm}
\begin{itemize}
\item[Note:] $\bar{v}$ denotes the mean, $v_\mathrm{med}$ the median, $v_{10}$
the $10^{\mathrm{th}}$ percentile, $v_\mathrm{max}$ the maximum, and
$\sigma_{v}$ the standard deviation of the horizontal plasma velocities.
\end{itemize}}
\end{center}
\end{table}
%-------------------------------------------------------------------------------

Immediately when the two polarities appeared on the surface, they started to 
move apart, reaching maximum separation, and shortly after began to approach 
each other. The branches of the two polarities have multiple barbs at their 
inner sides, indicating coalescence of same-polarity features. The small 
negative-polarity branch in the center represents the third pore, which spread 
out while decaying . The maximum separation is 45.6~Mm on July~20 at 01:36~UT. 
We computed the separation velocity for four evolutionary stages during the 
growth phase of the region. We tagged these four stages as (a) emerging phase 
15:30\,--\,17:30~UT on July~17, (b) maximum phase 08:00\,--\,10:00~UT on 
July~18, (c) second peak phase 00:00\,--\,02:00~UT on July~19, and (d) decaying 
phase 21:30\,--\,23:30~UT on July~19. The stages are indicated by arrows in 
Fig.~\ref{FIG04}, the four horizontal lines in Fig.~\ref{FIG05}, and are 
explained in detail in Sect.~\ref{SEC03.2}.

In stage (a) the separation rate is 0.26~km~s$^{-1}$, which is the highest of 
the four stages and signals the rise of magnetic flux tubes to the solar 
surface. High separation rates are expected \citep{Otsuji2011}. In the growth 
phase (b) the polarities are separating with 0.22~km~s$^{-1}$. In stage (c) the 
separation rate decreased to 0.07~km~s$^{-1}$. Toward the last stage (d), the 
separation rate reaches its lowest value of 0.02~km~s$^{-1}$, which marks the 
point when the separation of the polarities stops. The overall separation rate 
follows the trend discussed in \citet{Brants1985a}, \citet{Otsuji2011}, and 
\citet{Toriumi2012}: fast separation in the beginning and then slowing down 
while the region expands. Typical separation rates presented in these studies 
cover the range of 0.3\,--\,2.4~km~s$^{-1}$. The highest separation rate for the 
EFR in active region NOAA~12118 lies below this range. This EFR does not follow 
the trend that the separation rate is higher for flux tubes containing less 
magnetic flux, as \citet{Otsuji2011} concluded from a statistical study of 101 
EFRs. Finally, another peculiarity concerns the almost identical separation and 
shrinking rates as the active region grows and decays. This leaves the 
impression that a connected flux system rises and, starting at 02:30~UT on 
July~20, submerges as a whole.

%-------------------------------------------------------------------------------
%    Figure 8 : Frequency distributions 
%-------------------------------------------------------------------------------
\begin{figure*}[t]
\includegraphics[width=0.33\textwidth]{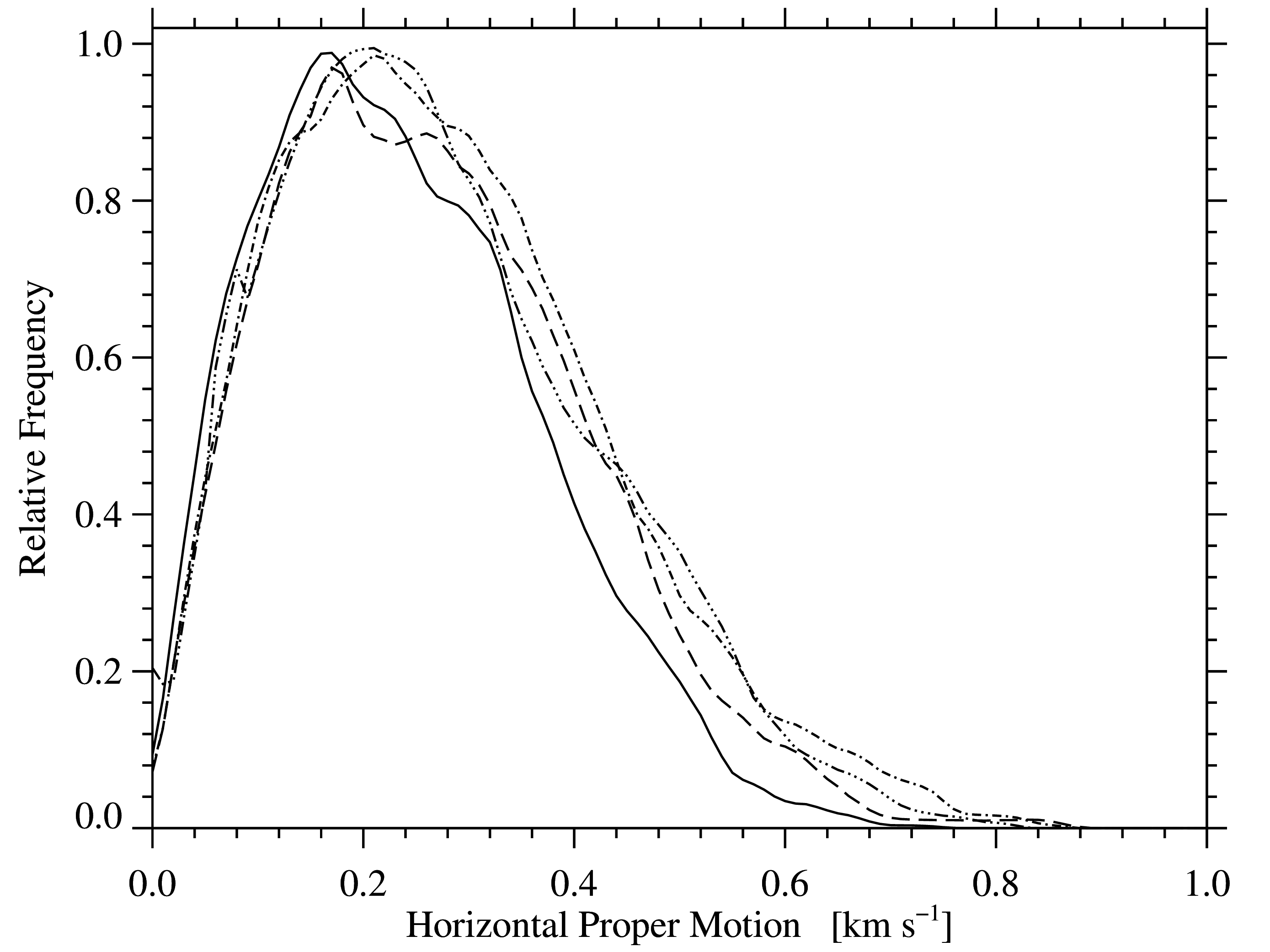}
\includegraphics[width=0.33\textwidth]{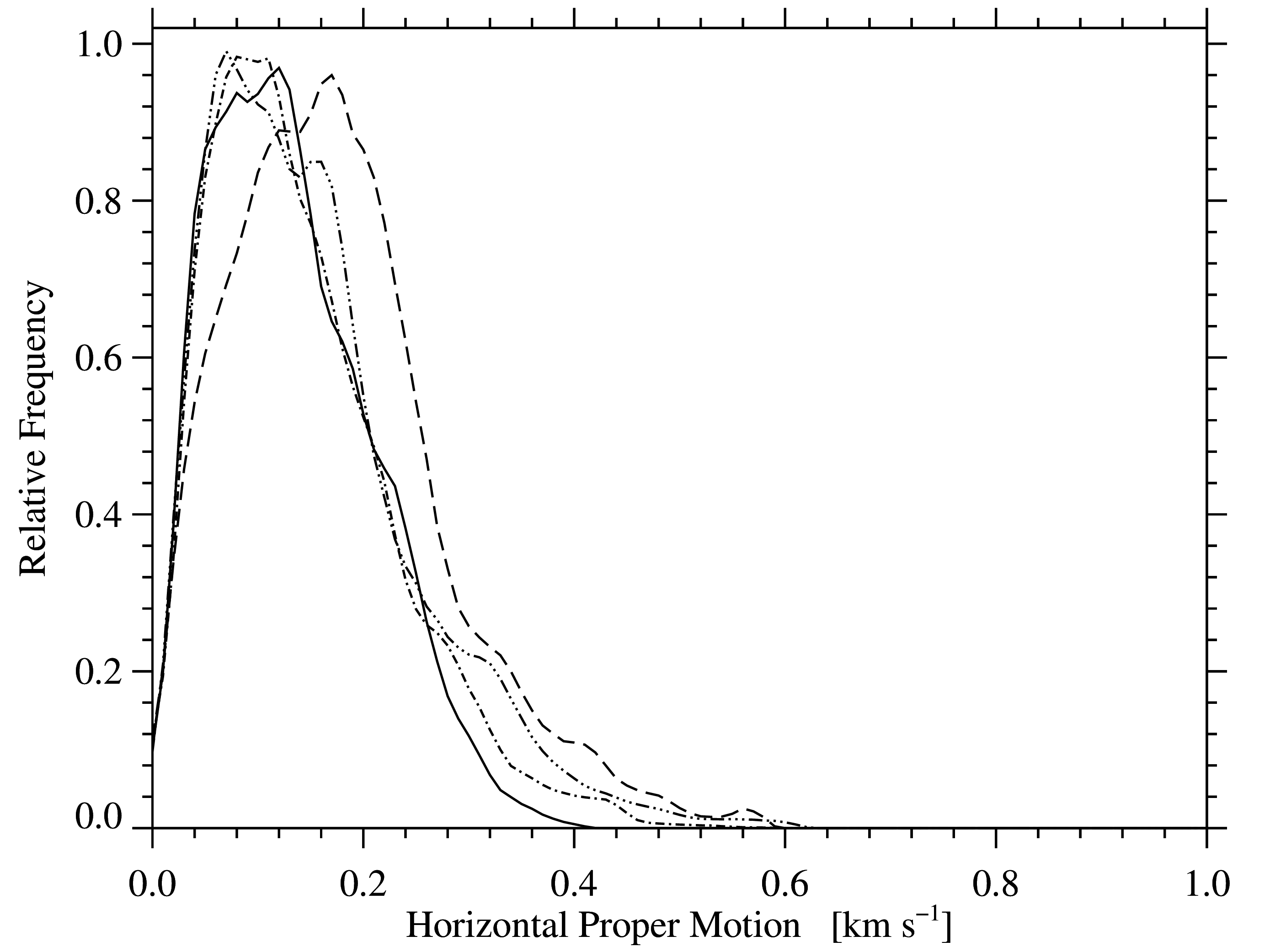}
\includegraphics[width=0.33\textwidth]{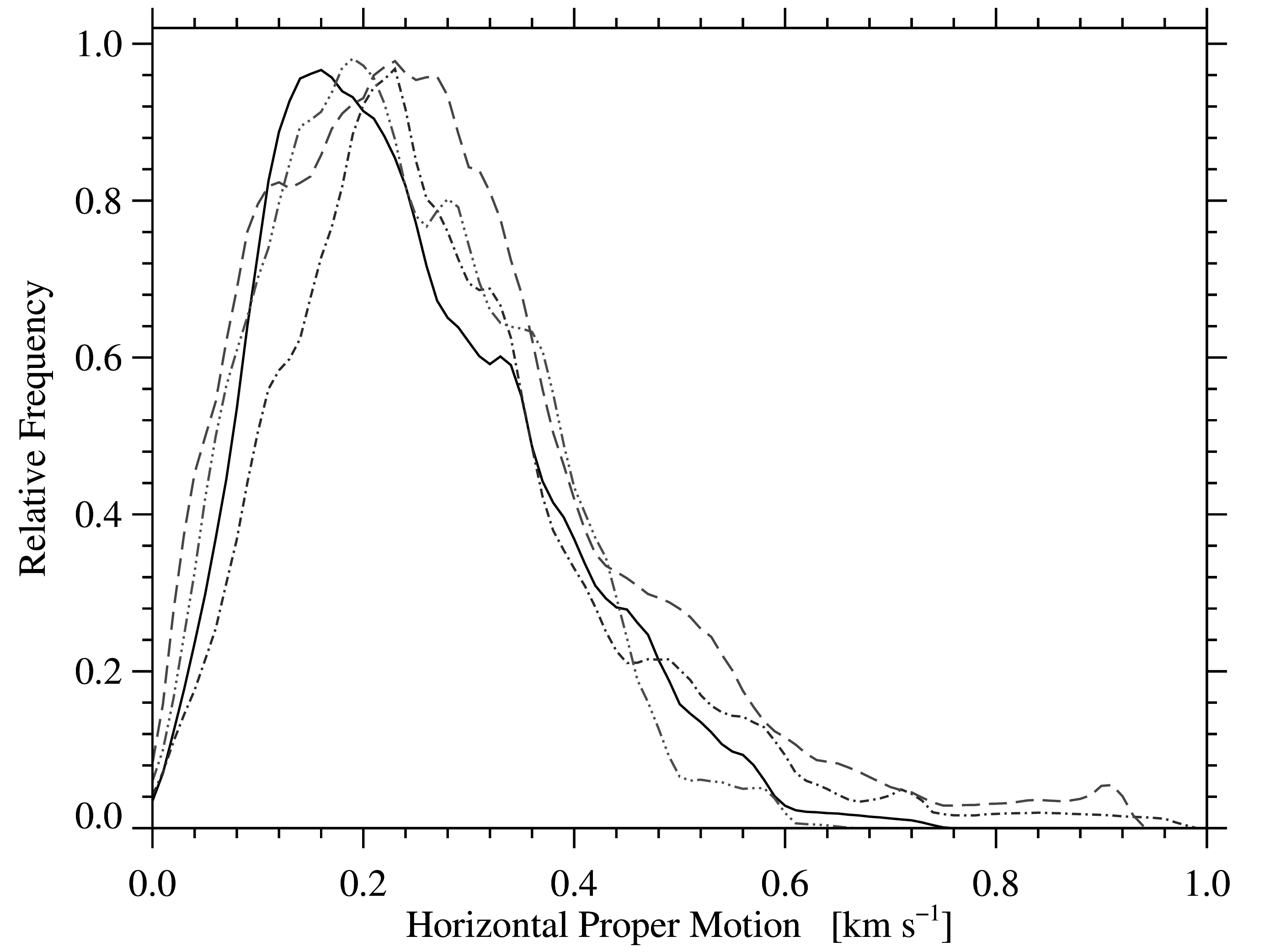}
\caption{Relative frequency distributions of horizontal proper motions for
    granulation (\textit{left}), pores (\textit{middle}), and granulation in the
    central region between the two major pores of opposite polarity
    (\textit{right}). The distributions for LCT panels (1)\,--\,(4) are depicted
    as solid, dash-dotted, dash-dot-dotted, and long dashed lines,
    respectively.}
\label{FIG08}
\end{figure*}
%-------------------------------------------------------------------------------

\subsection{Magnetic horizontal plasma velocity\label{SEC03.2}}

We used time-series of HMI magnetograms as input to compute horizontal plasma 
velocities with DAVE for four evolutionary stages of the region. These stages 
are marked in Figs.~\ref{FIG04} and~\ref{FIG05} by black arrows labeled a  to d. 
The averaged horizontal plasma  velocities are shown in Fig.~\ref{FIG06} as 
rainbow-colored arrows overlaid on time-averaged (two hours) magnetograms with a 
FOV of 80~Mm $\times$ 40~Mm. We created time-lapse movies to examine changes in 
the horizontal plasma velocities related to flux emergence in detail. 

In phase (a) when the active region is rising to the solar surface, we see 
strong motions along the horizontal central line pushing the two main patches of 
opposite polarities rapidly apart. The horizontal plasma  velocities are 
stronger for the negative-polarity region. The central part of the region, 
within $\pm 5$~Mm from the center coordinate (0~Mm, 0~Mm), is dominated by mixed 
polarities, and coherent motions are absent. Weak swirling motion patterns are 
present in the southern part of the region, for example, at coordinates (10~Mm, 
$-$10~Mm), as is evident in the time-averaged velocity vectors shown in 
Fig.~\ref{FIG06}.

%-------------------------------------------------------------------------------
%     Figure 09 : Zoom in view of exploding granule
%-------------------------------------------------------------------------------
\begin{figure*}[t]
\centering
\includegraphics[width=\textwidth]{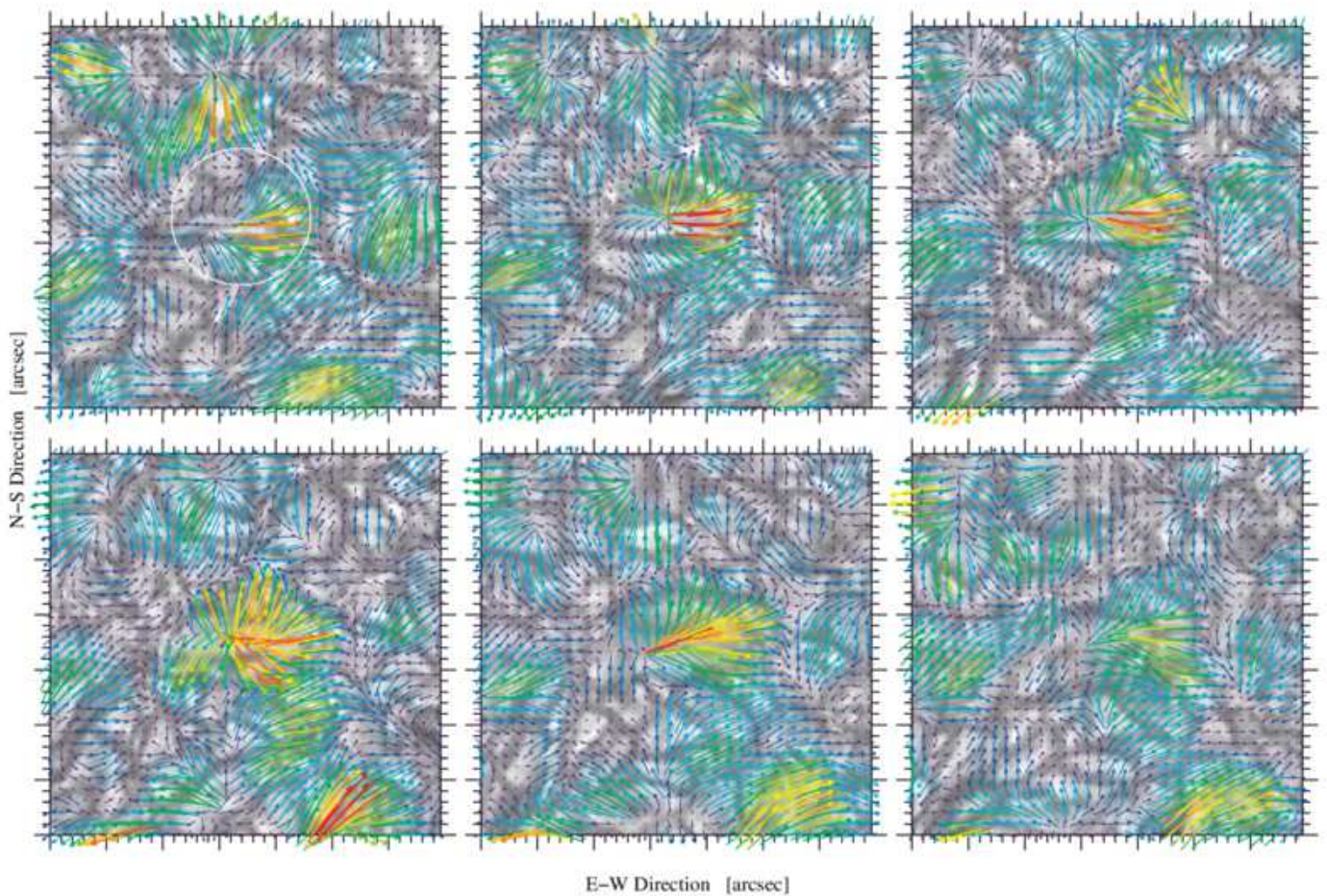}
\caption{Cutouts of 14\arcsec$\times$14\arcsec\ from blue continuum images 
    (\textit{right to left}) with a rapidly expanding granule in the center  at 
    six evolutionary stages from the second time-series starting at 08:36~UT 
    on July~18. Color-coded 5-minute-averaged velocity vectors (red indicates a 
    flow speed in excess of 0.7~km~s$^{-1}$) are superposed onto the first 
    image of the respective time-series. The white circle in the first panel 
    marks the location of the rapidly expanding granule.}    
\label{FIG10}
\end{figure*}
%-------------------------------------------------------------------------------

The region grows rapidly until July~18, when the fast expansion starts to slow 
down in phase (b). The magnetic flux in the two major pores \textsf{P1} and 
\textsf{N1} has become very compact. This flux concentration occurred about half 
a day later for the positive-polarity pore \textsf{P1}. We observe stronger 
proper motions at the borders of the main flux patches \textsf{P1} and 
\textsf{N1}. DAVE picks up the strongest outward motions to the east of the 
positive polarity \textsf{P1} and to the west of the negative polarity 
\textsf{N1}. These strong outward flows around major polarities are vaguely 
reminiscent of moat flows \citep{Sheeley1972, Meyer1974}, although they are not 
symmetric around the magnetic features, and penumbrae are absent. The dominant 
polarity patches also move apart as a whole. However, the unidirectional flow 
speed inside the confines of the pores is lower than the external horizontal 
plasma velocities. The mixed polarities in the active region are the locations 
of flux cancellation, whereas the dominant positive and negative polarity 
patches \textsf{P1} and \textsf{N1} are locations of flux coalescence, which is 
also noticeable in the space-time diagram depicted in Fig.~\ref{FIG05}. Compared 
to the bulk velocities of major flux concentrations (averaged over the magnetic 
area of the two major flux patches \textsf{P1} and \textsf{N1}), the mixed 
polarity patches have high horizontal plasma velocities, which are still 
multi-directional.

In phase (c) the separation of the two polarities slows down even further and 
almost comes to a halt. During this period, the negative polarity has started to 
disperse and is fragmented, while the positive polarity patch is more compact. 
However, both patches are now clearly separated, and the central part is devoid 
of any mixed polarities. The overall velocity pattern is patchy, but a strong 
eastward motion in the positive-polarity is still observed. The upper part of 
the negative-polarity patch still shows expanding motions. In general, the flow 
speed is lower than in phases (a) and (b).

Phase (d) marks the decay phase of the active region evolution. The magnetic 
area of both polarities is smaller and more fragmented. The negative polarities 
remain as small-scale flux elements. The horizontal plasma  velocities are 
reduced in the area covered by these small-scale magnetic elements, and the 
horizontal plasma  velocity pattern becomes more random for the more compact 
positive polarity patch. In general, the horizontal plasma velocities are no 
longer coherent, and the strongest velocities are found outside the active 
region.

%---- Table 3 ------------------------------------------------------------------
% \input{table3.tex}
\begin{table}[t]
\begin{center}
\caption{Parameters describing the horizontal proper motions computed from 
   averaging all four LCT maps (Fig.~\ref{FIG07}) for granulation 
   (\textit{top}), pores (\textit{middle}), and granulation in between the 
   pores \textsf{P1} and \textsf{N1} of opposite polarities 
   (\textit{bottom}).}\medskip
\small
\label{TAB03}
\begin{tabular}{ccccc}
\hline\hline
$\bar{v}$ & $v_\mathrm{med}$ & $v_{10}$ & $v_\mathrm{max}$ & $\sigma_{v}$\rule[-2mm]{0mm}{6mm}\\

[km~s$^{-1}$] & [km~s$^{-1}$] & [km~s$^{-1}$] & [km~s$^{-1}$] & [km~s$^{-1}$]\rule[-2mm]{0mm}{3mm}\\
\hline
$0.27$ & $0.26$ & $0.47$ & $0.83$ & $0.15$ \rule{0mm}{4mm}\\
$0.16$ & $0.14$ & $0.28$ & $0.55$ & $0.09$ \rule{0mm}{4mm}\\
$0.26$ & $0.24$ & $0.46$ & $0.83$ & $0.14$ \rule{0mm}{4mm}\\
\hline
\end{tabular}
\end{center}
\end{table}
%-------------------------------------------------------------------------------

To characterize velocities computed with DAVE, we present descriptive parameters 
in Table~\ref{TAB02}. The frequency distributions (not shown) for all four 
evolutionary stages are narrow with a high velocity tail. The mean velocity 
$\bar{v} \approx 0.10$~km~s$^{-1}$ is in agreement with the quiet-Sun values 
presented by \citet{Diercke2014} and \citet{Kuckein2016} for small-scale 
magnetic elements along an extended filament channel.

\subsection{Horizontal proper motions\label{SEC03.3}}

Using blue continuum images obtained with GREGOR/BIC, we examined the 
high-resolution spatial and temporal evolution of horizontal proper motions with 
LCT (Fig.~\ref{FIG07}). We have four time-series on 2014 July~18 starting at (1) 
07:56~UT, (2) 08:36~UT, (3) 09:13~UT, and (4) 10:09~UT. These times are centered 
on the active region at maximum growth. The FOV is rotated to match the 
orientation of the HMI continuum images and LOS magnetograms. The overall 
appearance of the LCT flow pattern is very different from those computed with 
DAVE. However, the LCT and DAVE maps cannot be compared because they sample
different flow fields.

In all four panels we have two large conglomerates of pores (\textsf{P1} and 
\textsf{N1} and one smaller pore \textsf{N2} at the top. In all four time-series 
the eastern pore \textsf{P1} is growing, whereas the western pore \textsf{N1} 
continuously decays. By the fourth time-series the smaller pore \textsf{N2} also 
begins to decay. The average area of about 82.2~Mm$^2$ and 34.3~Mm$^2$ for 
\textsf{P1} and \textsf{N1}, respectively, lies on the higher end of the values 
computed in the statistical study of pores by \citet{Verma2014}. Few flow 
patterns remain the same in all four panels. A recurring pattern is, for 
example, the central region between \textsf{P1} and \textsf{N1}, which contains 
abnormal granulation and is the site of many rapidly expanding granules. A 
moat-like outward motion encircles \textsf{N1} giving the impression that the 
pore is in the center of a larger supergranular cell. In addition, the dark 
features with strong magnetic fields have reduced LCT proper motions. 

Large rapidly expanding granules, which are often elongated, are present in all 
time-series across the entire FOV, but they are more frequent in the central 
part of the active region. Some of these granules reach a length of 6\arcsec\ 
and a width of 2\arcsec, whereas normal granules are smaller and more circular 
with diameters of around 2\arcsec\ \citep{Hirzberger1997}. The circle in 
Fig.~\ref{FIG07} marks the location of one such granule in close proximity to a 
magnetic bipole, which is indicative of an $\Omega$ loop rising through the 
photosphere. In the absence of magnetograms with high spatial resolution, it is 
difficult to find an alignment between the major axes of the elongated granules 
and the orientation of the magnetic field. Size, elongation, and location of 
this particular kind of granules suggest an association with emerging flux 
rather than pointing to extreme cases of regular exploding granules 
\citep{Namba1986, Rast1995, Hirzberger1999}. However, in both cases divergence 
centers mark the location of granules associated with flux emergence 
\citep{Ortiz2014} and regular exploding granules \citep[e.g.,][]{Bonet2005}. In 
consequence, we prefer the term `rapidly expanding' over `exploding' granules in 
this study related to flux emergence.

All pores contain multiple umbral dots \citep[cf.,][]{Sobotka2005} and around 
the two major pores finger-like structures intrude into the dark core 
\citep[cf.,][]{Scharmer2002, BellotRubio2008}, but the pores never developed 
penumbrae. In all four panels around 10\,--\,20 short-lived magnetic knots with 
a size smaller than one second of arc appear and disappear in the central part 
of the active region between \textsf{P1} and \textsf{N1}. The central part is 
characterized by horizontal channels of aligned intergranular lanes connecting 
the two magnetic polarities \citep[cf.,][]{Strous1995, Strous1999}. Both 
\textsf{P1} and \textsf{N1} exhibit inflows from the central part of the active 
region. In addition to the inflows, \textsf{N1} shows strong outward flows at 
its western periphery, in particular in panel~(2).

Panel~(2) shows very similar morphological and flow properties as panel~(1). 
Around \textsf{N1} the moat-like outward motion is stronger, with large granules 
fragmenting near its border. Horizontal alignment of granules extending between 
two polarities is present, but not as pronounced as in panel~(1). The most 
striking LCT flow feature in panel~(2) is the strong divergence region with flow 
vectors curling around the southwestern part of \textsf{P1}. This divergence 
pattern is strongest in panel~(2), but it is present in all panels. We marked it 
with a white rectangle in panel~(2). This diverging flow pattern suggests 
persistent upwelling and emergence of magnetic flux. 

In panel~(3) the overall flow pattern remains the same as in the previous panel. 
However, the magnitude of the velocity is lower. The inflows and outflows from 
the inner and outer parts of pores are still present. The strong divergence 
feature to the southwest of \textsf{P1} still exists. Various rapidly expanding 
granules are present, but they are not as large as in panels~(1) and (2). The 
LCT flow field in panel~(4) follows the flow pattern in the previous panels. In 
all four panels, the interior of pores does not have uniform and ordered flow 
vectors as seen in the DAVE maps. Although a few dominant flow features are the 
same in all panels, the flow field evolves significantly over half an hour. To 
quantitatively infer the properties of the central region with granulation, we 
computed relative frequency distributions for all four panels, which we discuss 
in Sect.~\ref{SECT03.4}. 

Similar to the space-time diagram of HMI magnetograms, we created a space-time 
diagram for 2.75~hours using four time-sequences (not shown). The computed 
separation speeds are for panels~(1) 0.61~km~s$^{-1}$, (2) 0.34~km~s$^{-1}$, (3) 
0.11~km~s$^{-1}$, and (4) 0.36~km~s$^{-1}$. These separation speeds are higher 
than the values derived from the HMI space-time diagram. The difference could 
arise from the fact that we tracked intricate details within pores in the blue 
continuum images, whereas in magnetograms we observe these pores as large 
polarity patches. Small changes in the high-resolution images can lead to large 
differences in the space-time diagrams.

\subsection{Characteristics of horizontal proper
    motions\label{SECT03.4}}

To gain insight into the statistical properties of the LCT velocities in 
different solar features we computed relative frequency distributions using 
binary masks. We depict these masks in the third panel of Fig.~\ref{FIG07}, 
where the contours mark pores and the central region with granulation. The 
respective distributions are compiled in Fig.~\ref{FIG08} and Table~\ref{TAB03}, 
which includes the various parameters of these distributions. The four 
distributions for granulation are broad, but without a high-velocity tail and 
with mean velocities extending between $\bar{v}=0.24$\,--\,0.28~km~s$^{-1}$. The 
corresponding standard deviation values are within the limits of $\sigma_v = 
0.13$\,--\,0.16~km~s$^{-1}$. These values for granulation reach only about half 
the numerical values presented by \citet{Verma2011} based on Hinode G-band 
images, but they are closer to the values $\bar{v}=0.35 \pm 0.21$~km~s$^{-1}$ of 
\citet{Verma2012b}.

The lower granular flow speed might result from the proximity to strong magnetic 
flux concentrations. The velocity distributions of the central region with 
granulation differs from the distribution for granulation per se. The 
distributions are narrower and have a high-velocity tail. However, the mean 
velocity remains roughly the same. The high-velocity tail is attributed to the 
presence of many rapidly expanding granules. The distributions of pores are 
narrow. The mean values cover the range between 
$\bar{v}=0.14$\,--\,0.18~km~s$^{-1}$ with standard deviation of $\sigma_v = 
0.07$\,--\,0.10~km~s$^{-1}$. Even though somewhat lower, the results agree with 
values computed by \citet{Verma2011, Verma2012b}. 

\subsection{Horizontal proper motions associated with rapidly expanding
    granules\label{SEC03.5}}

Taking advantage of the high spatial resolution provided by the GREGOR solar 
telescope, we  zoomed-in to scrutinize the fine details on the solar surface. 
One of the emerging active region's dynamical characteristics are large 
elongated granules \citep{Rezaei2012, Guglielmino2010}. Various stages of an 
expanding granule are presented in Fig.~\ref{FIG10}. The FOV of 14\arcsec\ 
$\times$ 14\arcsec\ is centered on the granule. The six panels are separated by 
5~min in time and thus cover a period of 30~min. The LCT velocities are averaged 
over 5~min. The divergence was also computed using the time-averaged velocity 
components (not shown).

The central region with granulation has a positive divergence from the 
beginning. However, the magnitude first increases with time before fading away 
and spreading after about 15~min. The strong divergence center is a well 
-observed property of exploding granules \citep[e.g.,][]{Hirzberger1999} as well 
as rapidly expanding granules in regions of flux emergence \citep{Rezaei2012}. 
The large size and aspect ratio in combination with the appearance of a dark 
central lane within the upwelling plasma favor the latter scenario. The mean 
divergence in the rapidly expanding granule (depicted by the white circle in 
Fig.~\ref{FIG10}) ranges from 2.0$\times 10^{-3}$\,--\,$6.3\times 
10^{-4}$~s$^{-1}$. The corresponding velocity vectors trace the full extent of 
the expanding granule. The strong divergence region is surrounded by a negative 
divergence ring, which closely follows the intergranular boundary. The length 
and width of the granule are about 3.6\arcsec\ and 1.6\arcsec, respectively, 
which agrees with previous observations \citep{Namba1986}. The mean velocity 
within the rapidly expanding granule for the first panel is 0.40~km~s$^{-1}$, 
which increases to 0.72~km~s$^{-1}$ for the fourth panel and decreases to 
0.34~km~s$^{-1}$ for the sixth panel, which is consistent with recent results of 
\citet{Palacios2012}. The highest velocity is encountered in the fifth panel and 
amounts to 1.25~km~s$^{-1}$. This value is much lower than the one of about 
4~km~s$^{-1}$ given by \citet{Guglielmino2010}, which can be easily explained by 
the very narrow sampling window (FWHM < 0.2\arcsec) employed by these authors, 
however. \citet{Verma2011} discussed the dependence of LCT results on sampling 
window properties and averaging times in detail.

%===============================================================================
%    Conclusions
%===============================================================================

\section{Summary and conclusions}

We presented one of the data sets taken during the early science phase of the 
GREGOR solar telescope. The high-resolution images of active region NOAA~12118 
captured with BIC allowed us to study the morphological and flow properties 
of a small active region. HMI continuum images and magnetograms provided 
valuable magnetic information. The major findings of this study can be summarized 
as follows. 

The evolution of NOAA~12118 started with the emergence of new flux on the solar 
surface followed by flux coalescence. In the emerging phase, many mixed-polarity 
features are in between the two major pores. This leads to a complex magnetic 
neutral line, which simplified with time. Even though the pores never developed 
penumbrae, we observed strong outward flows, which indicate the expansion of the 
active region when the pores push through the surrounding plasma. The 
alternative explanation that the outflows give the impression of a partial moat 
flow is less convincing in the absence of Evershed flow 
\citep[cf.,][]{VargasDominguez2008, VargasDominguez2010}, Evershed clouds 
\citep{CabreraSolana2006}, and moving magnetic features 
\citep[MMFs,][]{Harvey1973} as signatures of an extended radial flow system. 
Ultimately, high-resolution spectroscopic observations are needed to elucidate 
the origins of outflows. The growth rates for photometric area, magnetic area, 
and magnetic flux are about twice as high as the corresponding decay rates, 
which is consistent with previous studies \citep{Verma2012a}. The computed 
magnetic area is four to five times larger than the photometric area.

The HMI magnetograms were used to compute the horizontal plasma  velocities and 
blue continuum images of GREGOR provided the horizontal proper motions for the 
active region. However, the flow patterns are different and poorly correlated on 
account of several factors. First, the horizontal plasma velocities (DAVE) and 
plasma motions (LCT) track different, only weakly related physical processes 
\citep{Welsch2007}. Second, in the active region magnetic features are not 
advected by surface flows but rather by subsurface phenomena \citep{Zhao2010}. 
Third, the difference in the spatial resolution plays a significant role. The 
high-resolution of GREGOR images allow following flows due to motion and 
distortion of individual granules, which is not possible in HMI magnetograms.

The central region between the two major pores contained abnormal granulation 
and many large, elongated, and rapidly expanding granules. This central region 
also had different relative frequency distributions for granulation with 
high-velocity tails. The high-resolution image sequences facilitated precise LCT 
measurements of horizontal proper motions, and the results agree well with 
previous LCT studies \citep{Verma2011, Verma2012a, Verma2012b}. Even the 
intricate details of the flow fields belonging to a large, rapidly expanding 
granule are well captured by the LCT. Using the space-time diagram, we inferred 
the separation rates of the two major pores. The expansion and shrinking of the 
active region exhibit symmetries with respect to separation velocities (about 
$\pm 0.25$~km~s$^{-1}$) and length of the evolutionary phases (five days). The 
separation rates computed agree with previous work. 

The leaf-like structure that traced both polarities in the space-time diagram 
indicates an atypical behavior of active regions, where flux dispersal commonly 
characterizes the decay process. We speculate that the morphology of main pores 
is the reason because they never developed penumbrae and as a consequence never 
produced an Evershed flow \citep{Evershed1909}. The lack of a fully developed 
penumbra results in the absence of a uniform moat flow. In addition, the 
surroundings of the pores are void of any MMFs \citep{Harvey1973}, which play a 
key role in the decay of an active region by magnetic flux erosion. The region 
was also exceptional because the trailing part remained more stable than the 
leading part, which disintegrated first. Hence, the temporal evolution of this 
active region is consistent with the emergence of monolithic flux tube and its 
subsequent submergence, which leads to the characteristic leaf-like structure in 
the space-time diagram.

The evolution of the active region presented in this study closely follows the 
evolutionary path outlined by \citet{Otsuji2011}, \citet{Centeno2012}, and 
\citet{Toriumi2012}. In the present study, the two main pores have high 
separation rates in the beginning. Once the region reaches its maximum flux and 
size by the end of July~18, the separation rate slows down. This trend closely 
resembles the observation of an EFR by \citet{Toriumi2012} and is consistent 
with numerical simulations by \citet{Stein2011}. Similar to observations of 
\citet{Centeno2012}, the region in our study has elongated rapidly expanding 
granules appearing between the two main polarities, along with flux 
cancellation. These features are also seen in the simulation by 
\citet{Stein2011}, where the emerging magnetic field distorts the granules, 
stretching them in the direction of the horizontal field. In numerical 
simulations of \citet{Cheung2010} the horizontal expansion of magnetic plasma is 
prevalent while rising. The emergence of small magnetic features is followed by 
their coalescence. We also noted the same pattern, where the initial appearance 
of the magnetic features is followed by the growth of pores and their 
coalescence. In the simulated active region the mean flow pattern was radially 
outward, which we only observed at the leading and to a lesser extent at the 
trailing edges of the active region, whereas inflows into pores were sometimes 
observed in the interior of the active region.

In the future, we will combine spectropolarimetric observations of the GFPI and 
the GREGOR Infrared Spectrograph \citep[GRIS,][]{Collados2012}. High-resolution 
imaging along with precision infrared spectropolarimetry will allow us to follow 
the evolution of flows as well as the magnetic evolution of active regions from 
the photosphere to the chromosphere. Simultaneous observations with the Hinode 
\citep{Kosugi2007} and the Interface Region Imaging Spectrograph 
\citep[IRIS,][]{DePontieu2014} will complement the ground-based observations, 
along with synoptic data from SDO. Detailed information of flows and magnetic 
fields during the emerging phase are still needed to better constrain numerical 
simulations \citep{Rempel2014}.

%===============================================================================
%    Acknowledgements
%===============================================================================

\begin{acknowledgements}
The 1.5-meter GREGOR solar telescope was build by a German consortium under the 
leadership of the Kiepenheuer-Institut f\"ur Sonnenphysik in Freiburg with the 
Leibniz-Institut f\"ur Astrophysik Potsdam, the Institut f\"ur Astrophysik 
G\"ottingen, and the Max-Planck-Institut f\"ur Sonnensystemforschung in 
G\"ottingen as partners, and with contributions by the Instituto de 
Astrof\'{\i}sica de Canarias and the Astronomical Institute of the Academy of 
Sciences of the Czech Republic. SDO HMI and AIA data are provided by the Joint 
Science Operations Center -- Science Data Processing. MS is supported by the 
Czech Science Foundation under the grant 14-0338S. CD is supported by the German 
Science Foundation (DFG) under grant DE 787/3-1. This study is supported by the 
European Commission's FP7 Capacities Programme under the Grant Agreement number 
312495. SJGM is grateful for financial support from the Leibniz Graduate School 
for Quantitative Spectroscopy in Astrophysics, a joint project of AIP and the 
Institute of Physics and Astronomy of the University of Potsdam (UP).
\end{acknowledgements}

%===============================================================================
%    Bibliography
%===============================================================================
% \input{articlebibliography.tex}

%===============================================================================

\end{document}